# Gauging Public Acceptance of Conditionally Automated Vehicles in the United States

**Antonios Saravanos \*, Eleftheria K. Pissadaki, Wayne S. Singh, Donatella Delfino**

Division of Applied Undergraduate Studies, New York University, New York, NY 10012, USA;
\*  Correspondence: saravanos@nyu.edu; Tel.: +1-212-992-8725

**Abstract:** Public acceptance of conditionally automated vehicles is a crucial step in the realization of smart cities. Prior research in Europe has shown that the factors of hedonic motivation, social influence, and performance expectancy, in decreasing order of importance, influence acceptance. Moreover, a generally positive acceptance of the technology was reported. However, there is a lack of information regarding the public acceptance of conditionally automated vehicles in the United States. In this study, we carried out a web-based experiment where participants were provided information regarding the technology and then completed a questionnaire on their perceptions. The collected data was analyzed using PLS-SEM to examine the factors that may lead to public acceptance of the technology in the United States. Our findings showed that social influence, performance expectancy, effort expectancy, hedonic motivation, and facilitating conditions determine conditionally automated vehicle acceptance. Additionally, certain factors were found to influence the perception of how useful the technology is, the effort required to use it, and the facilitating conditions for its use. By integrating the insights gained from this study, stakeholders can better facilitate the adoption of autonomous vehicle technology, contributing to safer, more efficient, and user-friendly transportation systems in the future that help realize the vision of the smart city.

## 1. Introduction

The concept of a smart city is intrinsically linked with the integration of intelligent vehicles [1–3], which are essential for transforming urban environments into more efficient, sustainable, and livable spaces. Indeed, we now see cities progressively incorporating self-driving or automated vehicles (AVs) powered by artificial intelligence into their transportation mix, significantly impacting the architecture and environmental sustainability of urban infrastructure [4]. In this paper, we explore the public acceptance of a class of intelligent vehicles known as conditionally automated vehicles (CAVs), which have recently moved from human imagination to reality. This technology is already available for purchase, and is in use in some parts of the world (e.g., Germany [5] and the United States [6]). The Society of Automotive Engineers International (SAE) provides a definition of CAVs through their classification model, which prescribes six levels of automation for vehicles. The classification spectrum [7] starts at Level 0, which refers to vehicles that have no automation, and goes to Level 5 for the most advanced level of automation—full driving automation. CAVs adhere to Level 3 of this framework, as they have "automated driving features" that "can drive the vehicle under limited conditions and will not operate unless all required conditions are met" [7]. Hence, one is "not driving when these automated driving features are engaged", but must be prepared to drive "when the feature requests" [7].

Technology adoption models and, in particular, the unified theory of acceptance and use of technology (UTAUT) line of models, have been shown to be influenced by culture. In particular, a meta-analysis by Faaeq et al. [8] illustrated the effects of culture on the UTAUT model, and a meta-analysis by Tamilmani et al. [9] illustrated the effects of culture on the UTAUT2 model. Moreover, Foroughi et al. [10] concluded that "despite AVs' potential advantages and concerns, it is still being critically discussed and evaluated". Therefore, even though the factors that influence the acceptance of CAVs have been studied for a European audience, there is no certainty that the findings are applicable in the context of the United States. Furthermore, given the transformative effect that artificial intelligence has been having around the world, insight into the public perception of CAVs would be advantageous not only to those in the automotive industry, but also to policy makers at the local, state, and federal levels. Accordingly, the following are our two main research questions:

RQ1. What factors may influence public acceptance of CAVs in the United States?
RQ2. Are there any differences in public acceptance of CAVs in the United States based on age, experience, or gender?

The remainder of this paper is structured as follows. Section 2 reviews relevant literature to provide a general understanding of the research topic. Section 3 outlines the materials and methods that were used for this study, including the model that was selected for this research, as well as the corresponding hypotheses, the experiment that was carried out to measure public perception of the technology, and the characteristics of our sample. Section 4 describes the process used to analyze the data, as well as the results from that analysis. Section 5 discusses the findings. Finally, Section 6 concludes the paper, presenting the implications that result from this work, as well as the study's limitations and suggestions for further development.

## 2. Background

AVs are regarded as a groundbreaking technology that is poised to influence the trajectory of urban transportation in the future [10]. CAVs hold the promise of significantly enhancing road safety by removing the need for humans to be directly involved in the driving process [11] as well as introduce efficiencies that help cultivate the emergence of the smart city [12,13]. There are two implementations of the technology that should be acknowledged, both by German manufacturers. The first is Mercedes-Benz's Drive Pilot system, which has received authorization for commercialization in both Europe and the United States [14]. This system was initially tested in Nevada in the United States at speeds of up to 40 miles per hour [15]. Cars featuring this system have been approved for sale in California [6], and according to Reuters [16], the California Department of Motor Vehicles has approved the automated driving system to operate on specified highways under specific conditions, without the need for active driver involvement. The second system is BMW's Personal Pilot L3, which will be available in some of the BMW 7 Series cars beginning in March 2024 [17].

Research indicates that, even with relatively low market penetration rates, CAVs deliver substantial advantages to road safety, leading to a significant reduction in traffic conflicts [11]. CAVs are anticipated to offer other benefits, such as reduced road accidents, decreased traffic congestion, and new mobility options [18]. However, whether the environmental impact of such technology is positive or negative is as yet unsettled, and further effort is still required to provide greater insight. Lehtonen et al. [19] found that a greater inclination to adopt CAVs correlates with a traveler's expectation to reduce their usage of public transportation and, to a smaller degree, active travel modes. Furthermore, they noted that individuals who travel using multiple modes of transport frequently rely on public and active travel, and are therefore more inclined to experience a shift (either an increase or a decrease) in their engagement with these forms of transportation. Their findings indicate that Level 3 AVs could present a hurdle to sustainability by potentially prompting those who currently use public transport and active travel modes to shift toward using personal AVs. This is reflected in another study by Lehtonen et al. [20], which determined that CAVs could significantly boost car travel once they are accessible. This expectation to travel more by AV is fueled mainly by the desire to engage in leisure activities while driving autonomously, the perceived value of the system, and the capability of AVs to fulfill unaddressed travel needs by simplifying the travel experience.

The work of O'Hern and Louis [18] explored the perception of readiness to adopt CAV technology as well as intentions to use it, through the use of the Technology Readiness Index. The study surveyed 384 Finnish participants and revealed that optimism about the technology was positively correlated with the intention to use CAVs, meaning that more optimistic individuals were more inclined to use them. Conversely, lower levels of insecurity about technology (i.e., higher levels of security with technology) were also associated with a stronger intention to use CAVs. Additionally, previous experience with CAVs had a positive impact on the intention to use them. Louw et al. [21] used data from an online survey of 18,631 car drivers from 17 countries, including eight European nations, to investigate the intention to use automated driving functions in four operational design domains: motorways, traffic jams, urban roads, and parking. They found that the intention to use automated driving functions was generally high, particularly for automated parking functions. Younger respondents (under 39), males, and those with previous advanced driving assistance systems experience showed the highest intention to use automated driving functions. There was a wide variation across countries, with respondents from countries with lower GDPs and higher road death rates showing a higher intention to use all automated driving functions. Conversely, countries with higher GDPs and lower road death rates showed lower intention to use automated driving functions. Accordingly, the study recommended that strategies concerning the development and deployment of CAVs should be tailored to different markets.

Other studies have explored the technology without using a technology acceptance model. For example, Schrauth et al. [22] looked at 5827 participants from France, Germany, Slovenia, Spain, Sweden, Australia, and the United States using an iterative evaluation approach based on the ISO 9241-210:2015 standard. Their findings show a generally positive acceptance of CAVs among different road user groups, although they note that not everything was positive, and there were notable concerns. Moreover, their work found that acceptance levels varied between groups, with vulnerable road users being less accepting of the technology compared to drivers of non-automated cars.

While the literature investigating the adoption of automated vehicles is growing, and there are several systematic reviews on the topic (e.g., Rahman and Thill [23] and Nordhoff et al. [24]), few studies have explored CAVs within the context of the extended unified theory of acceptance and use of technology (UTAUT2) model, which is considered to be the standard contemporary model to understand the acceptance of technology. We were able to identify two such studies. The first is by Foroughi et al. [10], whose analysis revealed five key factors affecting adoption: confidence in the technology, the enjoyment derived from its use, the impact of social factors, how well it fits with the user's needs, and the ease of use. Moreover, they report that compatibility enhances the relationship between the anticipated performance of AVs and the user's intention to utilize them. It is important to note that the authors did not explicitly state in which country or countries they collected their data, in spite of the fact that such adoption is influenced by culture (e.g., Edelmann et al. [25]).

The second study is by Nordhoff et al. [26], who investigated the public acceptance of conditionally automated (SAE Level 3) vehicles across eight countries in Europe. The authors conducted a survey based on the UTAUT2 instrument, which was completed by 9118 car drivers. The findings revealed that the factors of hedonic motivation (HM), social influence (SI), and performance expectancy (PE) were significant predictors of the intention to buy and use these cars, with almost a third (28.03%) of participants sharing that they planned to buy such a car once they were available. Moreover, the findings indicated that over two-thirds (71.06%) of respondents perceived these cars as easy to use. The study also queried participants on whether they would engage in other activities while piloting an AV, and almost half (41.85%) affirmed such an intent. Interestingly, about half were unwilling to engage in other activities, potentially indicating that not everyone fully trusts the technology to a point where they would feel comfortable allowing it to drive unsupervised. When given examples of such activities, almost half (44.76%) said that they would talk to fellow passengers, over a third (41.70%) said that they would observe the landscape, and almost one in five (17.06%) said they would work. An additional insight offered by the study was the positive influence of facilitating conditions (FC) on effort expectancy (EE) and HM. Similarly, SI positively influenced EE, FC, HM, and PE. Also, HM influenced EE. Lastly, the three moderators of age, experience, and gender had no impact on behavioral intentions (BI).

## 3. Materials and Methods

In this section, we provide an overview of the model employed in this study, along with the corresponding hypotheses. We then describe the data collection instrument and the experimental procedures implemented to gather the data. Additionally, we present the sociodemographic profile of our sample. Finally, we conclude this section by detailing the data analysis methods that were employed.

### 3.1. Development of the Hypotheses and Model

To investigate user perception and acceptance of Level 3 technology in the United States, we first identified a model that could be used for that purpose. We selected an adaptation of Venkatesh et al.'s UTAUT2 model proposed by Nordhoff et al. [26], which they applied in Europe to study the acceptance of Level 3 vehicles. We outline our proposed model and present the respective hypotheses. The model incorporates the five constructs from the original UTAUT2 model: PE, EE, SI, FC, and HM. PE reflects the perceived value that would be derived from using the technology. EE captures the effort that one perceives would be needed to use the technology. SI corresponds to the effect that others have on an individual's perception. FC reflects the support one perceives they will have for using the technology. Finally, HM represents the perceived enjoyment that would be derived by using the technology. The direct references to the literature for these definitions are presented in Table 1. Accordingly, the following hypotheses are proposed:

**H1.** *EE positively influences BI.*

**H2.** *FC positively influences BI.*

**H3.** *HM positively influences BI.*

**H4.** *PE positively influences BI.*

**H5.** *SI positively influences BI.*

**Table 1.** Definition of factors.

| Factor | Definition |
| --- | --- |
| Behavioral Intention (BI) | "refers to a person's subjective probability that he will perform some behavior" [27] |
| Effort Expectancy (EE) | "the degree of ease associated with consumers' use of technology" [28] |
| Facilitating Conditions (FC) | "consumers' perceptions of the resources and support available to perform a behavior" [28] |
| Hedonic Motivation (HM) | "the fun or pleasure derived from using a technology" [28] |

| Performance Expectancy (PE) | "the degree to which using a technology will provide benefits to consumers in performing certain activities" [28] |
| Social Influence (SI) | "the extent to which consumers perceive that important others (e.g., family and friends) believe they should use a particular technology" [28] |

The UTAUT2 model also includes three moderators: age, experience, and gender. Each of these is examined with respect to the aforementioned five variables (i.e., EE, FC, HM, PE, and SI). It should be noted that the experience construct is broken down to evaluate prior experience with the following nine technologies: automated emergency braking, forward collision warning, blind spot monitoring, drowsy driver detection, lane departure warning, lane keeping assistance, adaptive cruise control, parking assist, and self-parking assist.

**H6.** *Age, experience, and gender moderate the influence of EE, FC, HM, PE, and SI on BI.*

Finally, Nordhoff et al. [26] also proposed the addition of the following 10 interrelations between these UTAUT2 factors that were not originally included in Venkatesh et al.'s UTAUT2 model. They argue that there is limited understanding of how the constructs of the UTAUT2 model interact with each other within the scope of conditional automation [26]. They first propose a path between EE and PE, which they attribute to existing scholarly work investigating automated vehicle acceptance (i.e., Herrenkind et al. [29], Nordhoff et al. [30], Zhang et al. [31]). They also highlight that this work aligns with the findings that are present in the broader technology acceptance literature, pointing to the work of Adams et al. [32], Karahanna et al. [33], and finally Venkatesh and Davis [34]. Further, they propose paths between SI and the other four factors: EE, FC, PE, and HM. They base this proposal on the work of Acheampong and Cugurullo [35], who determined that subjective norms (akin to SI) are positively associated with the perceived advantages of automated vehicles and the ease of operating automated driving technology, as well as a positive link between subjective norms and perceived behavioral control [26]. Acheampong and Cugurullo [26] also draw upon the findings of Nordhoff et al. [30], who reported that SI has a beneficial impact on performance expectations, the presence of support conditions, and the enjoyment derived from the activity.

With respect to FC, a path to EE, HM, and PE was proposed. This rationale was supported by the research of Nordhoff et al. [30], who indicated that there is a scarcity of information concerning the connection between FC and EE or PE, as well as the HM derived from the use of technology [26]. Furthermore, they observe that this study (i.e., Nordhoff et al. [30]) identified positive impacts of FC on both EE and the HM people perceived they would derive, yet found no connection between FC and PE. However, they did not discover a significant relationship in their own research.

We subsequently looked at HM. Nordhoff et al. [26], who point out that, generally, research into technology acceptance has shown that PE positively influences perceived usefulness and ease of use in this domain. They reference the studies of Koenig-Lewis et al. [36], as well as Teo and Noyes [37]. Concerning HM, they report a positive effect on EE, pointing to the work of Nordhoff et al. [30], who observed that HM had a favorable influence on EE, although its impact was not significant on PE. Moreover, they share that this finding is similar to what was reported in previous research by Herrenkind et al. [29]. Correspondingly, the following hypotheses are therefore proposed:

**H7.** *EE positively influences PE.*

**H8.** *FC positively influences PE.*

**H9.** *FC positively influences EE.*

**H10.** *FC positively influences HM.*

**H11.** *HM positively influences PE.*

**H12.** *HM positively influences EE.*

**H13.** *SI positively influences PE.*

**H14.** *SI positively influences EE.*

**H15.** *SI positively influences FC.*

**H16.** *SI positively influences HM.*

### 3.2. Instrument, Data Collection, and Sociodemographic Profile of Sample

After obtaining their consent to participate (which was reviewed by New York University's Institutional Review Board under IRB-FY2022-6642), a web-based experiment was then carried out on a total of 358 participants from the United States, who were presented with an information sheet outlining the L3 technology. Subsequently, participants were asked a series of six demographic questions to capture their age, salary, education, employment, gender, and marital status. The survey also included 24 questions that were adapted from the work of Nordhoff et al. [26] (which were in turn based primarily on Venkatesh et al.'s [28] UTAUT2 instrument). These questions relied on a 7-point Likert scale, ranging from strongly disagree to strongly agree (i.e., strongly disagree, moderately disagree, somewhat disagree, neutral, neither agree nor disagree, somewhat agree, moderately agree, and strongly agree).

The sociodemographic profile of our sample presents a diverse yet predominantly middle-aged, middle-income, and well-educated demographic, with a significant majority being married and employed full-time. The gender distribution was nearly balanced, with females constituting 48.60% ($n = 174$) and males representing 51.40% ($n = 184$) of the participants. In terms of age, most of the sample was between 26 and 55 years old, accounting for 80.28% of the participants, indicating a predominantly adult demographic. Specifically, individuals aged 26–30 formed the largest age group at 24.30% ($n = 87$), followed by those aged 31–35 at 20.39% ($n = 73$). The salary ranges showed a concentration in the middle-income brackets, with 29.05% ($n = 104$) of the participants earning between USD 30,000 and USD 49,999, and 27.09% ($n = 97$) earning between USD 50,000 and USD 69,999. A significant portion, 18.44% ($n = 66$), reported earning USD 90,000 or more. Educational attainment varied, but a substantial majority of the sample held a bachelor's degree (62.01%; $n = 222$), and 21.79% ($n = 78$) had completed a master's degree. Only a small fraction had professional degrees (0.56%, $n = 2$) or doctoral degrees (0.28%, $n = 1$). Marital status was predominantly married at 77.09% ($n = 276$), with single (never married) participants making up 18.99% ($n = 68$). Separated, divorced, and widowed statuses were minimal, each constituting less than 3% of the sample. Employment data indicated that a vast majority were in full-time employment (88.27%, $n = 316$), with part-time employment making up 6.70% ($n = 24$). Those retired accounted for 2.51% ($n = 9$), while 1.68% ($n = 6$) were seeking work, and a single student represented a mere 0.28% ($n = 1$) of the sample. The sociodemographic variables of the sample are presented in greater detail in Figure 1.

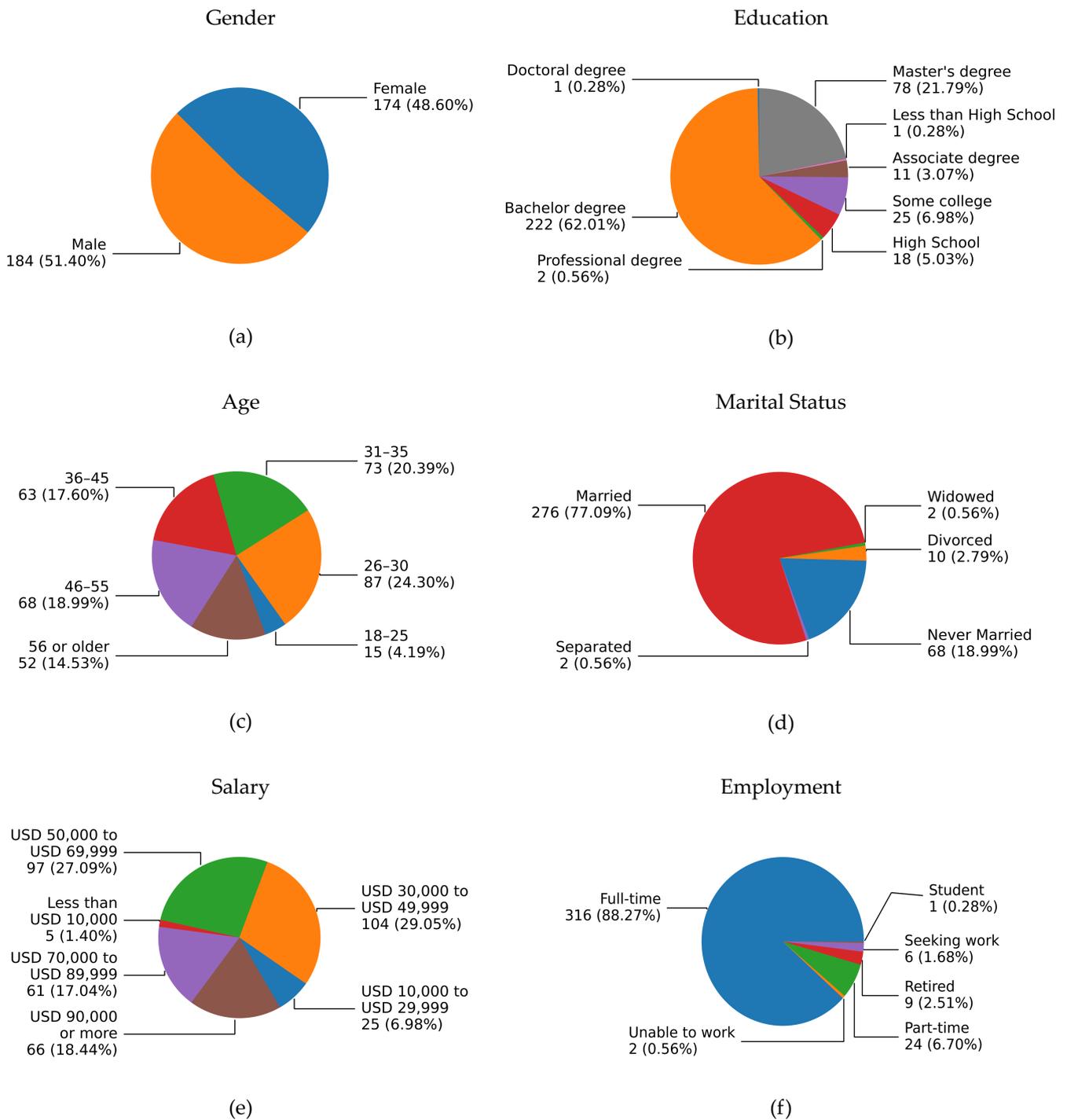

**Figure 1.** Pie charts illustrating sociodemographic variables: (**a**) gender; (**b**) education; (**c**) age; (**d**) marital status; (**e**) salary; (**f**) employment.

### 3.3. Data Analysis

To analyze the collected data, we relied on structural equation modeling (SEM). The approach is regarded as an extension of traditional linear modeling techniques [38]. It is recognized for its ability to work with complex models, and is "used to explain multiple statistical relationships simultaneously through visualization and model validation" [38]. There are two primary flavors of this approach; the first is covariance based (CB), and the other is partial least squares (PLS). Partial least squares structural equation modeling (PLS-SEM), is a method particularly recommended when working with complex models, data that is non-normal, and sample sizes that are small [39]. However, based on the guidelines put forth by Chin and Newsted [40], our sample size of 358 should be described as adequate and not small. Indeed, Pereira et al. [41] point to 100 as a threshold value, with samples lower than that being described as small, and samples greater than 100 being described as large. PLS-SEM relies on two elements: a measurement model that "specifies the relations between a construct and its observed indicators", and a structural model that "specifies the relationships between the constructs" [42].

## 4. Results

For our analysis, we applied PLS-SEM. A two-step approach (see Sarstedt et al. [43]) was taken to analyze the collected data. In the first instance, we evaluated the proposed measurement model and, subsequently, having established a reliable and valid measurement model, we tested the associated structural model. In the following sub-sections, we outline in greater detail both the steps that were taken and the respective results.

### 4.1. Measurement Model

In the first instance, we sought to assess reliability. First, we looked at indicator reliability, to establish that the manifest variables are properly associated. All of our factor loadings were above 0.708, which is the recommended minimum value (see Hair et al. [44]). Furthermore, we would note that the corresponding $t$-statistics were statistically significant to the 0.01 level ($p < 0.01$); in other words, the respective loadings were not due to chance. These findings are summarized in Table 2.

**Table 2.** Indicator reliability.

| Factor | Item | Loading | $t$-Statistic |
|--------|------|---------|---------------|
| BI | BI1 | 0.901 | 79.921 * |
| | BI2 | 0.822 | 30.404 * |
| | BI3 | 0.818 | 30.770 * |
| | BI4 | 0.879 | 54.441 * |
| | BI5 | 0.876 | 60.069 * |
| EE | EE1 | 0.884 | 64.781 * |
| | EE2 | 0.840 | 35.557 * |
| | EE3 | 0.870 | 52.995 * |
| FC | FC1 | 0.722 | 12.592 * |
| | FC2 | 0.764 | 23.882 * |
| | FC3 | 0.799 | 30.599 * |
| | FC4 | 0.711 | 14.744 * |
| HM | HM1 | 0.881 | 48.141 * |
| | HM2 | 0.865 | 38.603 * |
| | HM3 | 0.887 | 61.355 * |
| PE | PE1 | 0.802 | 25.477 * |
| | PE2 | 0.829 | 33.571 * |
| | PE3 | 0.809 | 32.128 * |
| | PE4 | 0.853 | 42.323 * |
| | PE5 | 0.874 | 53.694 * |
| SI | SI1 | 0.898 | 73.199 * |
| | SI2 | 0.894 | 55.268 * |
| | SI3 | 0.905 | 67.324 * |
| | SI4 | 0.863 | 62.717 * |

* $p < 0.01$.

We then evaluated the reliability of the latent construct. Common measures include composite reliability (CR) and Cronbach's alpha. We proceeded to test for each of the factors (i.e., BI, EE, FC, HM, PE, and SI). The BI factor was tested using five items and achieved excellent reliability with a Cronbach's alpha score of 0.911 and a CR score of 0.934. These values signify a very high level of consistency in responses, pointing toward the factor's robust internal consistency and reliability. Three items were used to test the EE factor. It demonstrated high reliability, with a Cronbach's alpha value of 0.832 and a CR value of 0.899. This indicates a strong internal consistency within the items measured. Tested using four items, the FC factor presented the lowest Cronbach's alpha value of 0.740 and a CR value of 0.837 among all factors. Although these values indicate acceptable reliability, they are comparatively lower, which may suggest a lesser degree of consistency compared to other factors. Three items were used for the HM factor, yielding high reliability, as shown by a Cronbach's alpha score of 0.851 and a CR score of 0.910. This suggests that the factor exhibits very high internal consistency and reliability. The PE factor showed high reliability scores, with a Cronbach's alpha value of 0.890 and a CR value of 0.919. The factor was tested using five items, and the results suggest a high degree of internal consistency. Finally, the SI factor was tested with four items, showing excellent reliability and internal consistency, as indicated by a Cronbach's alpha value of 0.913 and a CR score of 0.938.

In conclusion, all factors demonstrated good to excellent reliability, with both the Cronbach's alpha (ranging from 0.740 to 0.913) and CR (ranging from 0.837 to 0.938) values exceeding the widely accepted threshold of 0.7. Such high reliability scores validate the testing method and suggest that the measures used are appropriate and consistent. These values are outlined in Table 3.

**Table 3.** Construct reliability.

| Factor | Number of Items | Cronbach's Alpha | Composite Reliability |
|--------|-----------------|------------------|-----------------------|
| BI | 5 | 0.911 | 0.934 |
| EE | 3 | 0.832 | 0.899 |
| FC | 4 | 0.740 | 0.837 |
| HM | 3 | 0.851 | 0.910 |
| PE | 5 | 0.890 | 0.919 |
| SI | 4 | 0.913 | 0.938 |

We then moved on to evaluating validity. First we looked at convergent validity, relying on the average variance extracted (AVE) measure. Given that all of our AVE values, outlined in Table 4, were above the prescribed 0.5 threshold (see Hair et al. [44] for guidelines), we felt confident to proceed.

**Table 4.** Convergent validity.

| Factor | Number of Items | AVE |
|--------|-----------------|-----|
| BI | 5 | 0.739 |
| EE | 3 | 0.748 |
| FC | 4 | 0.562 |
| HM | 3 | 0.770 |
| PE | 5 | 0.695 |
| SI | 4 | 0.792 |

We then looked at discriminant validity, using the Fornell–Larcker criterion, which is outlined in Table 5. This helps us to assess whether a construct is distinct from other constructs in the model. For this evaluation, we compared the square root of the AVE for each construct with the correlation between constructs. The square root of the AVE should be larger than the correlations for adequate discriminant validity. This was indeed the case.

**Table 5.** Discriminant validity.

| Factor | BI | EE | FC | HM | PE | SI |
|--------|-------|-------|-------|-------|-------|-------|
| BI | **0.860** | | | | | |
| EE | 0.597 | **0.865** | | | | |
| FC | 0.438 | 0.623 | **0.750** | | | |
| HM | 0.772 | 0.606 | 0.514 | **0.878** | | |
| PE | 0.809 | 0.645 | 0.528 | 0.818 | **0.834** | |
| SI | 0.859 | 0.477 | 0.337 | 0.667 | 0.750 | **0.890** |

Note: The bolded values represent the square roots of the AVE.

### 4.2. Structural Model

Our investigation of the relationships between the factors, which involved evaluating the structural model and reporting on each independent variable, is summarized in Table 6. To start, BI explained 88% of the variance ($R^2 = 0.881$; adjusted-$R^2 = 0.851$). Such an $R^2$ can be interpreted as substantial, as per the guidelines of Chin [45] (i.e., ≥0.67) and the guidelines of Hair et al. [46] (i.e., ≥0.75). We then looked at each of the factors that we hypothesized to be contributors. As direct effects may not necessarily be comprehensive enough, we also looked at the total effect. From our findings, the construct SI stands out prominently in the results. It boasts both a substantial direct effect on BI ($\beta = 0.566$; $p < 0.01$), as well as a total effect ($\beta = 0.814$; $p < 0.01$). In terms of magnitude of the total effect, SI was followed by PE, which had a statistically significant direct and total effects ($\beta = 0.255$; $p < 0.05$). Respectively, this was followed by the EE factor, which did not have a statistically significant direct effect ($\beta = 0.121$; $p > 0.05$), but did have a statistically significant total effect ($\beta = 0.160$; $p < 0.05$). Likewise, HM did not have a statistically significant direct effect ($\beta = -0.008$; $p > 0.05$) but did have a statistically significant total effect ($\beta = 0.155$; $p < 0.05$). The smallest role was played by the FC factor, which also did not boast a statistically significant direct effect ($\beta = 0.011$; $p > 0.05$), but did have a statistically significant total effect ($\beta = 0.150$; $p < 0.01$).

**Table 6.** $R^2$ and adjusted-$R^2$.

| Factor | $R^2$ | Adjusted-$R^2$ |
|--------|-------|----------------|
| BI | 0.881 | 0.851 |
| EE | 0.510 | 0.506 |
| FC | 0.114 | 0.111 |
| HM | 0.538 | 0.535 |
| PE | 0.773 | 0.771 |

The second independent variable, EE, explained 51.0% of the variance, with an $R^2$ of 0.510 and an adjusted-$R^2$ of 0.506. Again, relying on the standards established by Chin [45] as well as Hair et al. [46], these findings can be described as moderate. In particular, with respect to Chin [45], the $R^2$ value falls between 0.33 (inclusive) and less than 0.67, and with respect to Hair et al. [46], the $R^2$ value falls between 0.50 (inclusive) and less than 0.75. The most prominent effect was played by FC, which had both a statistically significant direct ($\beta = 0.425$; $p < 0.01$) and total ($\beta = 0.522$; $p < 0.01$) effects. This was followed by HM, which had a similarly statistically significant direct and total effect ($\beta = 0.298$; $p < 0.01$), and then by SI, which likewise had statistically significant direct ($\beta = 0.135$; $p < 0.05$) and total ($\beta = 0.477$; $p < 0.01$) effects.

The next independent variable in our model was FC, which had an $R^2$ of 0.114 and an adjusted $R^2$ of 0.111. These values were below the classification of both Chin [45] and Hair et al. [46]. We bring attention to Chin [45], who described values between 0.19 (inclusive) and less than 0.33 as weak, and also Hair et al. [46], who described values between 0.25 (inclusive) and less than 0.50 as weak. Irrespectively, for completeness, we report that SI had both a statistically significant direct and total effect ($\beta = 0.337$; $p < 0.01$).

The fourth independent variable was HM. It had an $R^2$ of 0.538 and an adjusted-$R^2$ of 0.535 which, under the framework of both Chin [45] as well as Hair et al. [46], can be described as moderate. The greatest effect on the HM factor was played by SI, which had both a statistically significant direct ($\beta = 0.556$; $p < 0.01$) and total ($\beta = 0.665$; $p < 0.01$) effect. This was followed by FC, which was also statistically significant with respect to both direct and indirect effects ($\beta = 0.326$; $p < 0.01$).

The last independent variable in our model, PE, and had an $R^2$ of 0.773 and an adjusted-$R^2$ of 0.771 which, under both Chin [45] and Hair et al. [46], can be interpreted as substantial. The factors that were considered were, firstly, HM, which had a statistically significant direct effect ($\beta = 0.449$; $p < 0.01$) and a statistically significant total effect ($\beta = 0.495$; $p < 0.01$). Next, SI had both a statistically significant direct effect ($\beta = 0.349$; $p < 0.01$) and total effect ($\beta = 0.750$; $p < 0.01$). EE also had both statistically significant direct and total effects ($\beta = 0.154$; $p < 0.05$), while FC did not have a statistically significant direct effect ($\beta = 0.083$; $p > 0.05$), but did have a statistically significant total effect ($\beta = 0.310$; $p < 0.01$). The results are summarized in Table 7.

**Table 7.** Structural model results.

| Path | β (Direct) | t-Statistic (Direct) | β (Total) | t-Statistic (Total) |
|------|-----------|---------------------|-----------|---------------------|
| EE → BI | 0.121 | 1.711 | 0.160 | 2.098 * |
| EE → PE | 0.154 | 2.559 * | 0.154 | 2.559 * |
| FC → BI | 0.011 | 0.178 | 0.150 | 2.588 ** |
| FC → EE | 0.425 | 5.705 ** | 0.522 | 8.684 ** |
| FC → HM | 0.326 | 6.363 ** | 0.326 | 6.363 ** |
| FC → PE | 0.083 | 1.709 | 0.310 | 5.573 ** |
| HM → BI | -0.008 | 0.082 | 0.155 | 1.968 * |
| HM → EE | 0.298 | 3.093 ** | 0.298 | 3.093 ** |
| HM → PE | 0.449 | 7.272 ** | 0.495 | 8.628 ** |
| PE → BI | 0.255 | 2.288 * | 0.255 | 2.288 * |
| SI → BI | 0.566 | 5.840 ** | 0.814 | 13.096 ** |
| SI → EE | 0.135 | 2.048 * | 0.477 | 7.960 ** |
| SI → FC | 0.337 | 5.201 ** | 0.337 | 5.201 ** |
| SI → HM | 0.556 | 10.628 ** | 0.665 | 15.601 ** |
| SI → PE | 0.349 | 7.599 ** | 0.750 | 22.893 ** |

* $p < 0.05$; ** $p < 0.01$.

As for the moderating effects of age, experience, and gender on the five constructs (i.e., EE, FC, HM, PE, and SI) toward BI. We found that age did not influence the role played by EE, FC, HM, PE, and SI on BI to accept AV technology, and neither did the nine dimensions of experience. These results are presented in greater detail in Tables 8 and 9, respectively. Conversely, we did find gender to influence the HM factor, with the strength of the relationship between HM and BI being more prominent for men than women ($\beta = 0.357$; $p < 0.01$). The results can be seen in Table 10.

**Table 8.** Moderating effects of age on Behavioral Intention.

| Path | β (Direct) | t-Statistic (Direct) | β (Total) | t-Statistic (Total) |
|------|-----------|---------------------|-----------|---------------------|
| AGE → BI | 0.027 | 0.982 | 0.027 | 0.982 |
| AGE x EE → BI | -0.039 | 0.720 | -0.039 | 0.720 |
| AGE x FC → BI | 0.000 | 0.011 | 0.000 | 0.011 |

| Path | β (Direct) | t-Statistic (Direct) | β (Total) | t-Statistic (Total) |
|---|---|---|---|---|
| AGE x HM → BI | 0.100 | 1.356 | 0.100 | 1.356 |
| AGE x PE → BI | 0.004 | 0.053 | 0.004 | 0.053 |
| AGE x SI → BI | -0.039 | 0.720 | -0.039 | 0.720 |

**Table 9.** Moderating effects of experience on Behavioral Intention.

| Path | β (Direct) | *t*-Statistic (Direct) | β (Total) | *t*-Statistic (Total) |
|---|---|---|---|---|
| EXP1 → BI | 0.162 | 0.843 | 0.162 | 0.843 |
| EXP1 x EE → BI | -0.042 | 0.145 | -0.042 | 0.145 |
| EXP1 x FC → BI | -0.177 | 0.686 | -0.177 | 0.686 |
| EXP1 x HM → BI | -0.042 | 0.110 | -0.042 | 0.110 |
| EXP1 x PE → BI | 0.468 | 1.089 | 0.468 | 1.089 |
| EXP1 x SI → BI | -0.218 | 0.605 | -0.218 | 0.605 |
| EXP2 → BI | -0.112 | 0.495 | -0.112 | 0.495 |
| EXP2 x EE → BI | -0.109 | 0.325 | -0.109 | 0.325 |
| EXP2 x FC → BI | -0.164 | 0.546 | -0.164 | 0.546 |
| EXP2 x HM → BI | 0.238 | 0.641 | 0.238 | 0.641 |
| EXP2 x PE → BI | 0.444 | 0.872 | 0.444 | 0.872 |
| EXP2 x SI → BI | -0.144 | 0.457 | -0.144 | 0.457 |
| EXP3 → BI | 0.082 | 0.462 | 0.082 | 0.462 |
| EXP3 x EE → BI | 0.178 | 0.789 | 0.178 | 0.789 |
| EXP3 x FC → BI | 0.020 | 0.082 | 0.020 | 0.082 |
| EXP3 x HM → BI | -0.372 | 1.016 | -0.372 | 1.016 |
| EXP3 x PE → BI | -0.235 | 0.630 | -0.235 | 0.630 |
| EXP3 x SI → BI | 0.340 | 0.911 | 0.340 | 0.911 |
| EXP4 → BI | 0.365 | 1.277 | 0.365 | 1.277 |
| EXP4 x EE → BI | -0.145 | 0.401 | -0.145 | 0.401 |
| EXP4 x FC → BI | 0.189 | 0.547 | 0.189 | 0.547 |
| EXP4 x HM → BI | -0.490 | 0.969 | -0.490 | 0.969 |
| EXP4 x PE → BI | 0.000 | 0.000 | 0.000 | 0.000 |
| EXP4 x SI → BI | -0.002 | 0.003 | -0.002 | 0.003 |
| EXP5 → BI | 0.043 | 0.206 | 0.043 | 0.206 |
| EXP5 x EE → BI | -0.003 | 0.009 | -0.003 | 0.009 |
| EXP5 x FC → BI | 0.143 | 0.482 | 0.143 | 0.482 |
| EXP5 x HM → BI | 0.386 | 0.866 | 0.386 | 0.866 |
| EXP5 x PE → BI | -0.686 | 1.337 | -0.686 | 1.337 |
| EXP5 x SI → BI | -0.048 | 0.112 | -0.048 | 0.112 |
| EXP6 → BI | 0.041 | 0.203 | 0.041 | 0.203 |
| EXP6 x EE → BI | 0.218 | 0.741 | 0.218 | 0.741 |
| EXP6 x FC → BI | 0.133 | 0.509 | 0.133 | 0.509 |
| EXP6 x HM → BI | -0.013 | 0.038 | -0.013 | 0.038 |
| EXP6 x PE → BI | -0.381 | 0.675 | -0.381 | 0.675 |
| EXP6 x SI → BI | -0.223 | 0.464 | -0.223 | 0.464 |
| EXP7 → BI | -0.089 | 0.575 | -0.089 | 0.575 |
| EXP7 x EE → BI | 0.087 | 0.315 | 0.087 | 0.315 |
| EXP7 x FC → BI | 0.123 | 0.494 | 0.123 | 0.494 |
| EXP7 x HM → BI | 0.364 | 1.174 | 0.364 | 1.174 |
| EXP7 x PE → BI | -0.325 | 0.816 | -0.325 | 0.816 |
| EXP7 x SI → BI | -0.065 | 0.491 | -0.065 | 0.491 |
| EXP8 → BI | -0.028 | 0.162 | -0.028 | 0.162 |
| EXP8 x EE → BI | -0.257 | 1.305 | -0.257 | 1.305 |
| EXP8 x FC → BI | 0.371 | 1.772 | 0.371 | 1.772 |

| | | | |
|---|---|---|---|
| EXP8 x HM → BI | -0.110 | 0.477 | -0.110 | 0.477 |
| EXP8 x PE → BI | -0.071 | 0.480 | -0.071 | 0.480 |
| EXP8 x SI → BI | -0.117 | 0.390 | -0.117 | 0.390 |
| EXP9 → BI | -0.235 | 0.714 | -0.235 | 0.714 |
| EXP9 x EE → BI | -0.036 | 0.126 | -0.036 | 0.126 |
| EXP9 x FC → BI | -0.070 | 0.149 | -0.070 | 0.149 |
| EXP9 x HM → BI | 0.681 | 1.153 | 0.681 | 1.153 |
| EXP9 x PE → BI | -0.124 | 0.178 | -0.124 | 0.178 |
| EXP9 x SI → BI | 0.043 | 0.206 | 0.043 | 0.206 |

**Table 10.** Moderating effects of gender on Behavioral Intention.

| Path | β (Direct) | *t*-Statistic (Direct) | β (Total) | *t*-Statistic (Total) |
|---|---|---|---|---|
| GENDER → BI | -0.018 | 0.330 | -0.018 | 0.330 |
| GENDER x EE → BI | -0.073 | 0.667 | -0.073 | 0.667 |
| GENDER x FC → BI | 0.044 | 0.468 | 0.044 | 0.468 |
| GENDER x HM → BI | 0.357 | 2.622 * | 0.357 | 2.622 * |
| GENDER x PE → BI | -0.225 | 1.517 | -0.225 | 1.517 |
| GENDER x SI → BI | -0.018 | 0.33 | -0.018 | 0.330 |

* $p < 0.01$.

## 5. Discussion

The introduction of hands-off automated driving technologies by automotive giants such as Mercedes-Benz [47] and BMW [17] marks a pivotal moment in the field of automotive engineering—a transformation that was once confined to the realm of human imagination. This advancement is a direct result of significant strides made in artificial intelligence and robotics. Mercedes-Benz has been at the forefront of this revolution with their Drive Pilot technology, designed to enable automated driving in traffic conditions up to a top speed limit of 60 km/h [48]. This innovation represents a fundamental shift in how we perceive and interact with automobiles. This semi-automated driving technology merges the boundaries between driver assistance and full automation, bringing us closer to a future where cars are not just vehicles, but intelligent companions on the road. Similarly, BMW's introduction of the Personal Pilot L3, especially in their 7 Series, scheduled for release in March 2024, underscores the rapid pace at which automotive technology is evolving. This system is expected to offer an unprecedented level of autonomy in driving, thereby setting new standards in the luxury car sector. It is not just about the automation of driving tasks, it is about redefining the experience of mobility.

These technological advancements are not isolated developments; instead, they are integral to the broader vision of creating smart cities. In a smart city, technology is seamlessly integrated into the urban fabric to enhance the efficiency of services and the quality of life for its inhabitants [12,13]. Automated vehicles, such as those developed by Mercedes-Benz and BMW, are crucial in this context. They offer more than just convenience since they have the potential to revolutionize urban transportation systems. By enhancing road safety, lowering the environmental impact of vehicles, and potentially circumstances reducing traffic congestion, these technologies contribute significantly to the sustainable development of urban spaces. However, it should be recognized that there is research (e.g., Lehtonen et al. [19] and Lehtonen et al. [20]) showing that the convenience of CAVs would most likely lead to an increase in the use of vehicles by individuals, resulting in a negative effect on the sustainable development of urban spaces.

In essence, these technologies are not just changing how we drive, they are reimagining the very fabric of urban living, promising a future that is safer, more efficient, and more attuned to the needs of society. However, for any technology to be used, it must first find acceptance. To that end, we surveyed 358 individuals on their perceptions of such technology in the United States. Our study then applied PLS-SEM to evaluate the proposed measurement model and the associated structural model in the context of AV technology acceptance. The robustness of our measurement model is evidenced by indicator reliability, with all factor loadings exceeding the recommended minimum thresholds and displaying statistical significance, suggesting that the manifest variables are well-associated. The construct reliability was also confirmed, with Cronbach's alpha and CR values well above the accepted benchmark of 0.7 for all constructs, indicating high internal consistency.

Regarding our first research question, which concerned the factors that may influence the public acceptance of CAVs in the United States, the structural model revealed the varying impact of different constructs on the behavioral intention to accept AV technology. Notably, SI emerged as a significant predictor, underscoring the profound effect of social factors on technology adoption. Its direct and total effects on behavioral intention were substantial, as were those for PE, indicating its pivotal role in shaping behavioral intentions. Conversely, EE, HM, and FC, despite showing significant total effects on behavioral intention, did not present significant direct effects on behavioral intention, suggesting that they may exert influence through other constructs rather than directly affecting behavioral intention. These findings align with the multi-faceted nature of technology acceptance, where various factors interplay to shape user attitudes and intentions. Attention should also be drawn to the differences between these findings and those of Nordhoff et al. [26], who explored the public reception of CAVs in Europe. Their study found that the perceived HM derived from the use of the technology has the greatest influence on acceptance, followed by SI (i.e., what others think) and PE. Moreover, EE and FC did not play a role in acceptance.

Our study's findings, wherein SI is the primary acceptance factor, could indicate a more community-oriented or peer-influenced approach to technology acceptance, indicating a strong dependence on societal norms and peer opinions. This stands in contrast to the findings of Nordhoff et al. [26], where the prominence of HM and EE indicated a more individualistic and user-centric approach in European markets, which suggests a culture where personal satisfaction and ease of use are key factors for acceptance. The prominence of HM in the European sample also suggests that users are increasingly looking for enjoyable and satisfying experiences.

While PE remains important in both studies, its relative position is lower in the European sample. This difference could indicate a market that has evolved to expect high performance as a given, thus bringing the focus more on ease of use. The rise in the importance of EE suggests a growing user preference for intuitive and user-friendly systems. The dissimilarity in importance of FC in the current study could be interpreted in several ways. It might indicate an improvement in general technological infrastructure, making this factor less of a concern for users, or it could reflect a change in the type of systems or products evaluated, whereby infrastructure is less critical to the user experience. The lesser importance of FC in the European study could be due to the differences in technological infrastructure or higher baseline expectations for technology in Europe. This might suggest that European users have shifted their focus from infrastructure availability to the user experience and ease of use.

These findings highlight the need for culturally sensitive approaches in system design and marketing strategies. What appeals to users in the United States might not resonate as strongly with European users, and vice versa. For example, Edelmann et al. [25] point to the work of Atchley et al. [49], who note that, in the United States, driving a car is deeply ingrained in the culture and linked to the notion of personal liberty, as well as a moderate tolerance for risk. Consequently, American drivers aim to reach their destinations swiftly, but they do not prioritize speed above everything else. The varying user preferences between the United States and Europe underscore the importance of regional customization in product development. Developers and marketers should tailor their approaches based on the specific characteristics and preferences of each market. For companies operating globally, these variations point to the necessity of a diversified strategy.

Understanding regional differences in user behavior and preferences is key to developing products and services that are successful across different markets. These findings also stress the need for broader market research that accounts for cultural and regional differences; which is crucial for businesses aiming to expand their reach and appeal to diverse user bases across the globe.

Our work found that a comparable number of respondents believed Level 3 cars to be easy to use (87% for our sample vs. 71.06% in the Nordhoff et al. [26] European sample). There was, however, a difference in that United States participants are more likely to buy such a car (69% in our sample vs. 28.03% in the Nordhoff et al. [26] European sample).

Regarding our second research question, exploring whether there are any differences in public acceptance of CAVs in the United States based on age, experience, or gender, we found only a moderating effect for gender, with HM having a more prominent influence on intention to accept for men. Given that other moderators did not have a statistically significant effect, this finding could be a result of the technology not being complex, but rather easy to use, regardless of the moderating factors explored (i.e., age, experience, and gender).

## 6. Conclusions

This research explores a component of intelligent urban design by examining the elements that shape the public's readiness to embrace CAVs, specifically SAE Level 3, within the United States. This study contributes to the existing body of knowledge by providing empirical evidence on the determinants of technology acceptance, and by offering a comprehensive analysis of the factors that drive the behavioral intention to use AVs. An adapted version of the UTAUT2 model and corresponding instrument was utilized for this purpose. An experimental study was conducted with 358 United States participants, who were provided with a descriptive scenario detailing Level 3 technology. In addition to gathering demographic data through a set of questions, participants were queried on their views regarding the technology. Data analysis was performed using PLS-SEM. Findings indicate that SI is the most significant factor affecting technology acceptance, followed by PE, EE, HM, and FC. Our findings indicate that the acceptance of AV technology is heavily influenced by the perceived benefits and the social endorsement of the technology. SI was found to have a strong direct and total effect on BI, suggesting that individuals' acceptance is significantly swayed by the opinions and behaviors of others within their social circle. Similarly, PE was a critical determinant, with its significant effect underscoring the importance of the perceived efficiency and effectiveness of the technology in influencing user acceptance.

Additionally, HM, SI, EE, and FC were all found to positively impact the perceived usefulness of the technology (i.e., PE), while FC, HM, and SI positively affected perceived ease of use (i.e., EE). SI was also seen to affect HM, as well as FC, positively. It was noted that 87% of participants found a Level 3 car to be user-friendly, and 69% expressed an intention to purchase such a car once available. Regarding secondary activities during non-driving time, 30% would converse with fellow passengers, 24% would browse the internet or watch videos/TV, 36% would enjoy the scenery, 34% would rest or relax, 32% would eat or drink, 24% would socialize with friends or family, 28% would work, 21% would read, 27% would attend to children, and 13% would play games. The analysis of moderating effects provided additional depth, showing that while age and experience did not significantly alter the relationships between the constructs and behavioral intention, gender played a role, with HM being a stronger predictor of adoption intention among men. The gender differences identified in the influence of HM on BI highlight the necessity of acknowledging and addressing the diverse motivations across different user groups. This underscores the need for a nuanced approach to user segmentation and targeting in strategy formulation. This insight is crucial for tailoring communication and marketing strategies to different segments of the potential user base. In the remainder of this section, we go over the significance of these findings by presenting the theoretical and practical implications of this work. We then conclude the section by highlighting the limitations of our efforts and presenting future research directions.

### 6.1. Theoretical, Practical, and Policy Implications

Our findings have implications that extend beyond academic discourse, offering tangible strategies for those engaged in the development and retail of such vehicles as well as policy-makers (i.e., local, state, and federal governments). By integrating the insights gained from this study, stakeholders can better facilitate the adoption of AV technology in the future, contributing to safer, more efficient, and user-friendly transportation systems that help realize the vision of the smart city. With respect to theory, our work illustrates that Nordhoff et al.'s [26] adaptation of the UTAUT2 model can indeed be effectively applied to investigate user acceptance within the United States cultural context. Concurrently, it also highlights the differences between the two cultural contexts, informing adoption theory as a whole.

Looking to industry, the findings indicate that, for the successful implementation and widespread acceptance of AV technology, efforts should be directed not only toward enhancing the performance features of AVs, but also toward leveraging SI and addressing the different motivational factors that affect various user groups. This follows the high explanatory power of the SI factor concerning the intention to accept the technology. Accordingly, companies could leverage social proof and endorsements to promote adoption. In terms of design, our findings suggest that designers should prioritize features that enhance the user experience and meet performance expectations. Considering the gender differences in HM, design elements should be inclusive and cater to a diverse user base, with features that are universally appealing and others that are tailored to different demographic groups. Similarly, marketing and communication strategies could be tailored to highlight positive testimonials and endorsements from early adopters. The gender-specific influence of HM suggests that marketing campaigns should be gender-sensitive, potentially highlighting different features or benefits that appeal more to men or women.

For policymakers, the insights from this study emphasize the importance of addressing public perceptions and expectations regarding AVs. Campaigns can be designed that encourage those who are successfully using the technology to share their experiences with those around them.

### 6.2. Limitations and Future Research Directions

This study, while comprehensive in its approach, is not without limitations, although it should be acknowledged that such limitations simultaneously inform the design and direction of subsequent research efforts. The primary limitation was the cross-sectional nature of its design, which allowed us to examine user perceptions of the technology only at a specific point in time. Hence, it would follow that a richer sample that includes perceptions over a longer period of time would offer a stronger understanding of how such perceptions of this technology evolve. Another limitation concerns the maturity and availability of the technology that was studied (i.e., Level 3 cars) which, at the time of data collection, was not available for purchase in the United States. Accordingly, our experiment relied on an information sheet outlining the CAV technology, which was provided to stimulate the imagination of the participants. It is important to stress that different results might be obtained if participants had the opportunity to use the technology before being surveyed. As this technology diffuses and becomes more widespread, a follow-up study to confirm findings would be of value to our community. A third limitation concerns the use of self-reported measures, which can be subject to response biases. Consequently, future research might benefit from incorporating objective data where possible to corroborate self-reported perceptions. For our analysis, we relied on PLS-SEM, which is a common approach used in similarly designed studies (e.g., [26,50–54]). However, it should be recognized that complementary analyses to SEM-PLS can be performed, such as AHP (e.g., [55]), fsQCA (e.g., [56]), and ANN (e.g., [57]), that could potentially add further insight to our findings. Lastly, we should address our sample being collected exclusively in the United States. As we highlighted earlier in our work, perceptions of this technology and the practice of technology acceptance modeling vary by culture, suggesting the need for caution when extending these results to other settings. It would, therefore, be beneficial to investigate whether results from other cultures or geographical settings would align with our findings.

**Author Contributions:** Conceptualization, A.S.; methodology, A.S.; data curation, A.S.; writing—original draft preparation, A.S., E.K.P., W.S.S. and D.D.; writing—review and editing, A.S., E.K.P., W.S.S. and D.D.; funding acquisition, A.S. and D.D. All authors have read and agreed to the published version of the manuscript.

**Funding:** This research was funded in part by a New York University School of Professional Studies Dean's Research Grant.

**Institutional Review Board Statement:** The study was conducted in accordance with the Declaration of Helsinki and approved by the Institutional Review Board of New York University (protocol code IRB-FY2022-6642).

**Informed Consent Statement:** Informed consent was obtained from all subjects involved in the study.

**Data Availability Statement:** The data that support the findings of this study are available from the corresponding author, A.S., upon reasonable request.

**Conflicts of Interest:** The authors declare no conflicts of interest.